# Analysis of Gait-Event-related Brain Potentials During Instructed And Spontaneous Treadmill Walking - Technical Affordances and used Methods

**Cornelia Herbert[1*], Jan Nachtsheim[1,2], Michael Munz[2*]**

***Abstract***

To improve the understanding of human gait and to facilitate novel developments in gait rehabilitation, the neural correlates of human gait as measured by means of non-invasive electroencephalography (EEG) have been investigated recently. Particularly, gait-related event-related brain potentials (gERPs) may provide information about the functional role of cortical brain regions in human gait control. The purpose of this paper is to explore possible experimental and technical solutions for time-sensitive analysis of human gait-related ERPs during spontaneous and instructed treadmill walking. A solution (HW/SW) for synchronous recording of gait- and EEG data was developed, tested and piloted. The solution consists of a custom-made USB synchronization interface, a time-synchronization module and a data merging module, allowing temporal synchronization of recording devices for time-sensitive extraction of gait markers for analysis of gait-related ERPs and for the training of artificial neural networks. In the present manuscript, the hardware and software components were tested with the following devices: A treadmill with an integrated pressure plate for gait analysis (zebris FDM-T) and an Acticap non-wireless 32-channel EEG-system (Brain Products GmbH). The usability and validity of the developed solution was tested in a pilot study (n = 3 healthy participants, n=3 females, mean age = 22.75 years). Recorded EEG data was segmented and analyzed according to the detected gait markers for the analysis of gait-related ERPs. Finally, EEG periods were used to train a deep learning artificial neural network as classifier of gait phases. The results obtained in this pilot study, although preliminary, support the feasibility of the solution for the application of gait-related EEG analysis..

**Key words:** human gait analysis, machine learning, motor potentials, event-related potentials (ERPs), gait-related ERPs, cognition

## 1. Introduction

The rehabilitation of patients with gait-related health problems plays an important role in medical health care [1, 2]. Stroke, aging, overweight, medication, diseases such as diabetes mellitus II and mild cognitive impairments as well as other life circumstances such as accidents at work or at home can lead to impairments in human gait [3, 4]. In particular, subtle impairments in gait are often overseen. Problems in gait control are therefore not always diagnosed properly because the factors contributing to changes in human gait patterns are multi-factorial. While the physical symptoms of impaired gait can be diagnosed by physiotherapists, neuropsychologists and doctors with the help of well-validated behavioral tools and neuropsychological tests [5], the neural correlates of human gait are still not considered in sufficient detail in the rehabilitation routine. Lately, there has been a huge interest in the neural correlates of human locomotion [6]. Methodologically, the analysis of event-related brain potentials (ERPs) and the analysis of changes in oscillatory neural activity as measured via electroencephalography (EEG) have become most important non-invasive methods for investigating human locomotion including gait [7]. The EEG, the analysis

[1]: Department of Applied Emotion and Motivation Psychology, Institute of Psychology and Education, Ulm University, Germany

[2]: Department of Medical Engineering and Mechatronics, Ulm University of Applied Sciences, Ulm, Germany

* Correspondence: cornelia.herbert@uni-ulm.de, Michael.munz@thu.de

of ERPs in particular, allows the investigation of changes in brain activity related to certain external or internal sensory events in the time range of milliseconds. The electrophysiological correlates of arm-, finger-, or limb-movements, for example, are well investigated [8, 9, 10, 11, 12]. Likewise processes that are involved in movement control during motor imagery have been studied intensively [13,14]. Regarding human motion and movement control, two types of EEG measures are of particular interest. First, the modulation of event-related-potentials (ERPs, [7, 10]) and second, the modulation of changes in oscillatory activity such as measures related to event-related-(de)synchronization (ERDS/ERS) [16]. Whereas ERPs are phase-locked to a particular event of interest, event-related-(de)synchronization describes frequency-band specific changes in EEG rhythms appearing due to changes in externally or internally paced cognitive, affective and motor activity. Changes in ERD/ERS are well validated in the literature [16, 17]. However, studies investigating the modulation of gait-related event-related potentials (gERPs) are still underrepresented. Regarding ERP measurement during walking or running, previous studies were using dual task paradigms (e.g., solving a cognitive task (primary task)) while walking or running on a treadmill (secondary task). However, these studies aimed at analyzing ERPs to the primary task while simultaneously canceling out brain activity from the secondary task (e.g., walking, running) as a source of noise elicited by the movement of walking or running [10, 18, 19]. Research that looks at gait-specific ERP modulation elicited during spontaneous or instructed walking is still sparse. This research, however, is of relevance for clinical research interested in improving the diagnosis and treatment of gait disorders with a primarily neural origin (e.g., stroke patients, Parkinson disease) or secondary neural origin (e.g., amputees, traumatic injuries, mentally disabled patients, elderly people with risk of falls, or overweight people with diabetes type II disorder), to name but a few of the disorders, that can be accompanied by CNS-related cortical gait impairments. One of the main reasons for this lack of research about gait-specific neural correlates is due to technical reasons. Another reason concerns physiological movement-artifacts produced by the gait itself [19]. Yet, another reason concerns assumptions about the control of highly automated movement in the human brain. The analysis of gait-related ERPs (gERPs) requires - as does analysis of ERPs in general - precise timing and triggering of the events of interest. In the case of gERPs, technical solutions are needed that capture a) the different periods of the human gait cycle (for a graphical illustration of the human gait cycle see Figure 1) and b) the concomitant EEG activity, time-locked to the onset of the different periods of the gait cycle (e.g., initial contact, etc. …). The events need to be recorded with temporally highest precision (in the millisecond range). Importantly, the gait events need to be synchronized with the EEG recordings without loss of temporal information and without data errors due to jittering in the trigger signal caused by imprecise data interfacing or uncertainty in signal and data transfer from one device to the other. Second, any bodily movement produces movement artefacts in the EEG signal that can overlay the gait-event or signal of interest. Regarding gERPs, such movement artifacts are temporally confounded with the gait-event and the EEG signal of interest. For instance, the initial contact (IC), i.e. one of the most prominent events of the gait cycle, can produce movement artifacts when the feet touch the ground (see Figure 1). Such gait-related movement artifacts may, however, occur with limited variability when gait is controlled in the laboratory, for instance, during treadmill walking at constant pace (see Figure 3). Finally, human gait is highly automatic. However, this does not mean that the human gait cycle is not under the control of higher-order cortical brain structures such as cortical motor and attention networks. Studies have shown that even in conditions in which walking is highly autonomous (walking on a treadmill at constant speed and pace) and therefore, only under limited attentive control, it can interfere with the performance of a number of mental cognitive tasks [18, 20, 21, 22].

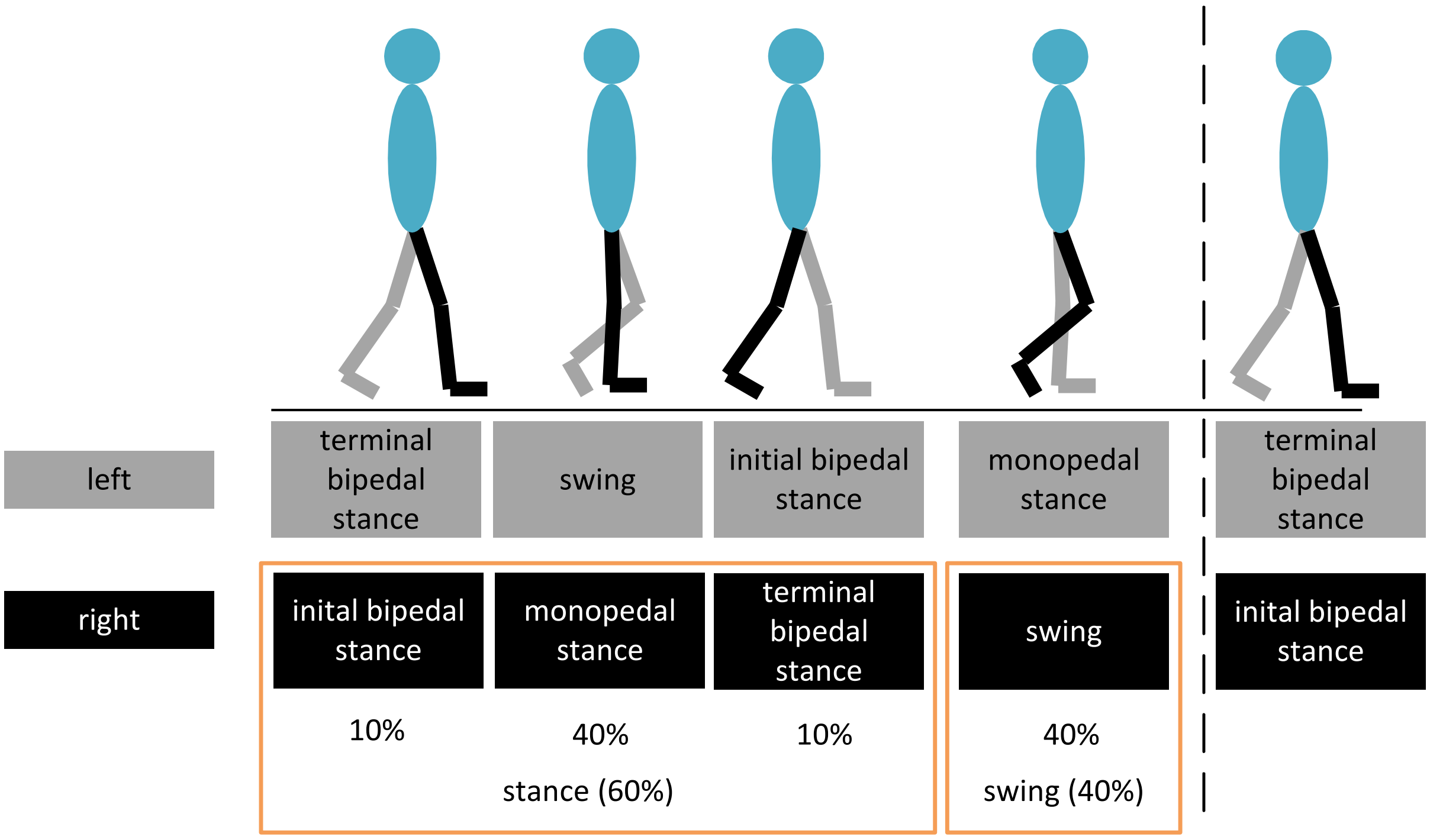

Figure 1: Monopedal (single-support) and bipedal (double support) phases of gait over one gait cycle. (modified after [23]), with corresponding average percentages of gait cycle. Each initial bipedal stance phase begins with initial contact (IC), followed by loading response (LR). Monopedal stance covers mid stance (MS), terminal bipedal stance begins with terminal stance (TS), and ends with toe off (TO), directly leading to the swing phase.

## 2. AIM OF THE PRESENT STUDY

As outlined above, studying the neural correlates of human gait is challenged by a number of technical restrictions. Therefore, the aim of the present study was to explore possible technical solutions for time-sensitive analysis of human gait-related ERPs (gERPs) during spontaneous and instruction guided treadmill walking. To realize these solutions, several hard- and software components are necessary for time-sensitive synchronous gait- and ERP-data interfacing and recording. In the present study, the technical solution chosen was tested and piloted in 3 healthy volunteers from whom continuous EEG-data was recorded from an electrically shielded high-density 32-channel EEG-system with active electrodes during uninstructed and instructed treadmill walking. Gait was recorded via a pressure plate implemented in the treadmill. Next, the gait cycle and corresponding gait markers were decoded following standardized classifications of human gait. Two key questions were explored:

1) Can gait-related event-related brain potentials (gERPs) be analyzed based on human gait markers decoded by force plates?
2) Can the analysis of gait-related ERPs be automated by applying machine learning approaches based on artificial neural networks (ANN)?

## 3. Materials and Methods

### *3.1 Structural Design and system overview*

Figure 2 provides an overview of the structural design including the technical devices and systems used in the study. The arrows between the different devices or systems illustrate the interfaces for data recording and data transfer established by the technical solution. The structural design used in the present study mirrors those of many previous EEG studies that analyze EEG during treadmill walking. The design comprises four independent systems whose data flow is combined and temporally synchronized. A) The stimulation unit which consists of a computer and the software for the presentation of stimuli and tasks (e.g., cognitive tasks, instructions, etc.) during treadmill walking. B) The recording units. These consist of two different computers, one being directly connected via a parallel port to the stimulation unit and used for timely precise EEG recordings, the other one being connected to the treadmill and pressure plate via USB cable and to the treadmill software for gait recording. In addition, in the present technical solution, a third device capturing data from inertial sensors (IMUs, for whole body motion capture) was implemented. The IMUs provide the possibility for an automatic analysis of gait events and patterns without the use of video-based systems. To integrate the IMUs, the trigger from the simulation software was split and send to a third recording device (notebook, personal computer) that was used for motion capture examined via inertial sensor systems. The participants receive the instructions sent from the stimulus PC. To this end, the stimulus PC is connected with a television screen located in front of the treadmill at the eye level of the participants. The instructions and stimuli are simultaneously sent via the stimulus presentation software to the EEG system and used as markers/triggers to later identify each walking condition and epochs of the gait events.

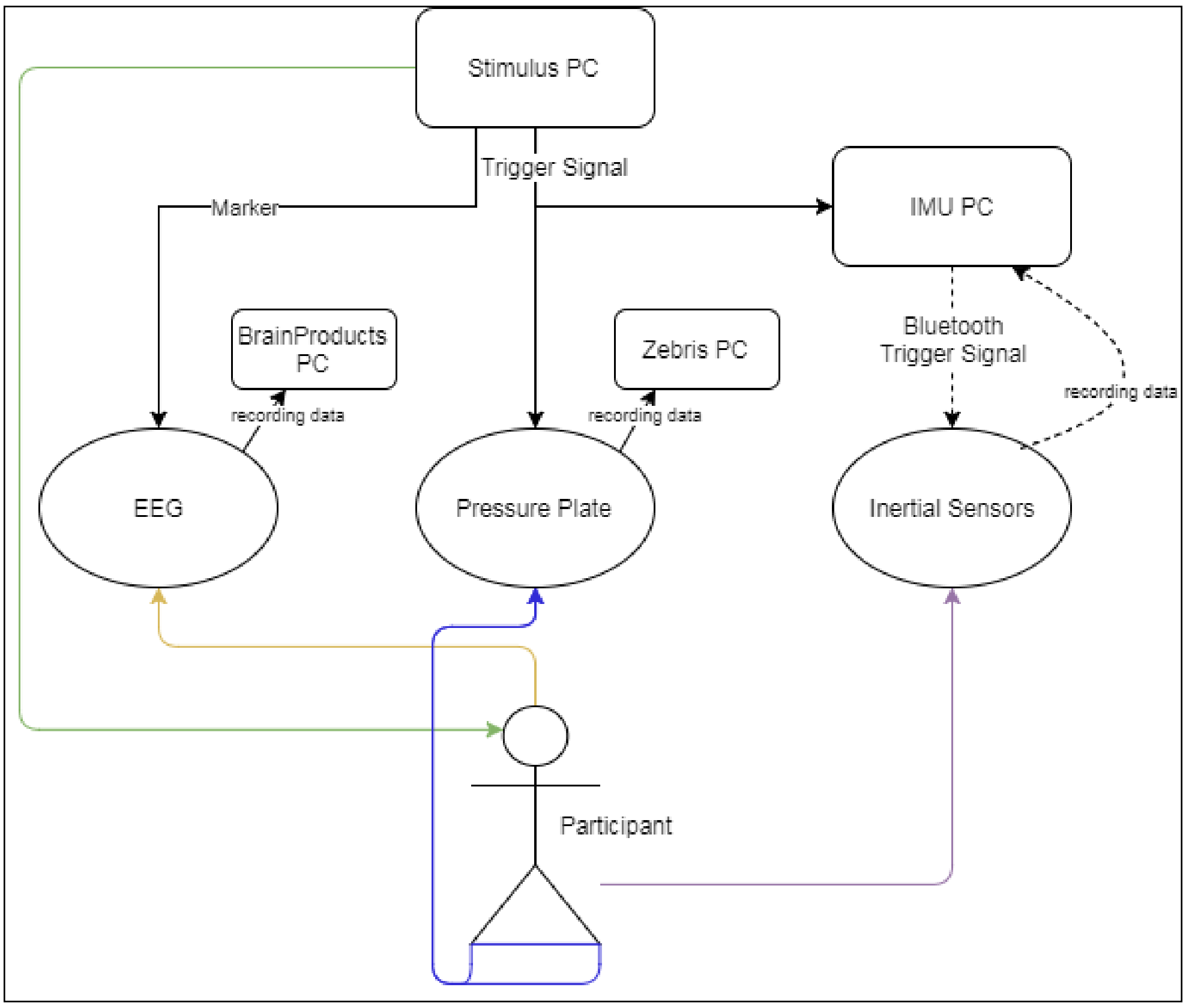

Figure 2: Structural design of the technical solution developed in this study. The solution includes the interfaces between the subsystems and recording devices. The stimulus PC can send triggers to the EEG, the pressure plate and the recording software of the inertial sensors (IMUs). Each system sends the recorded data to a corresponding PC.

*3.2 EEG system*

The EEG system is an active non-wireless high-density EEG system frequently used in scientific studies (Brain Products GmbH, Gilching, Germany). The actiCAP electrodes of the EEG system are connected with a multichannel BrainAmp amplifier and a computer with recording software (Brain Vision Recorder, Brain Products

GmbH, Gilching, Germany). The active electrode EEG system uses a hardware solution that leads to lower noise compared to passive electrodes. The EEG recording systems and EEG recording computer is connected via a linear parallel port (LPT) and a USB splitter box to the stimulus PC which allows precise triggering in the range of < 2ms without signal jittering. The stimulation PC is used for triggering to synchronize gait and EEG data (see Figure 2). The EEG was recorded with a sampling rate of 1000 Hz using the standard 10-20 channel montage for EEG recording. Impedance was controlled and kept constant across electrodes below 10 kOhm.

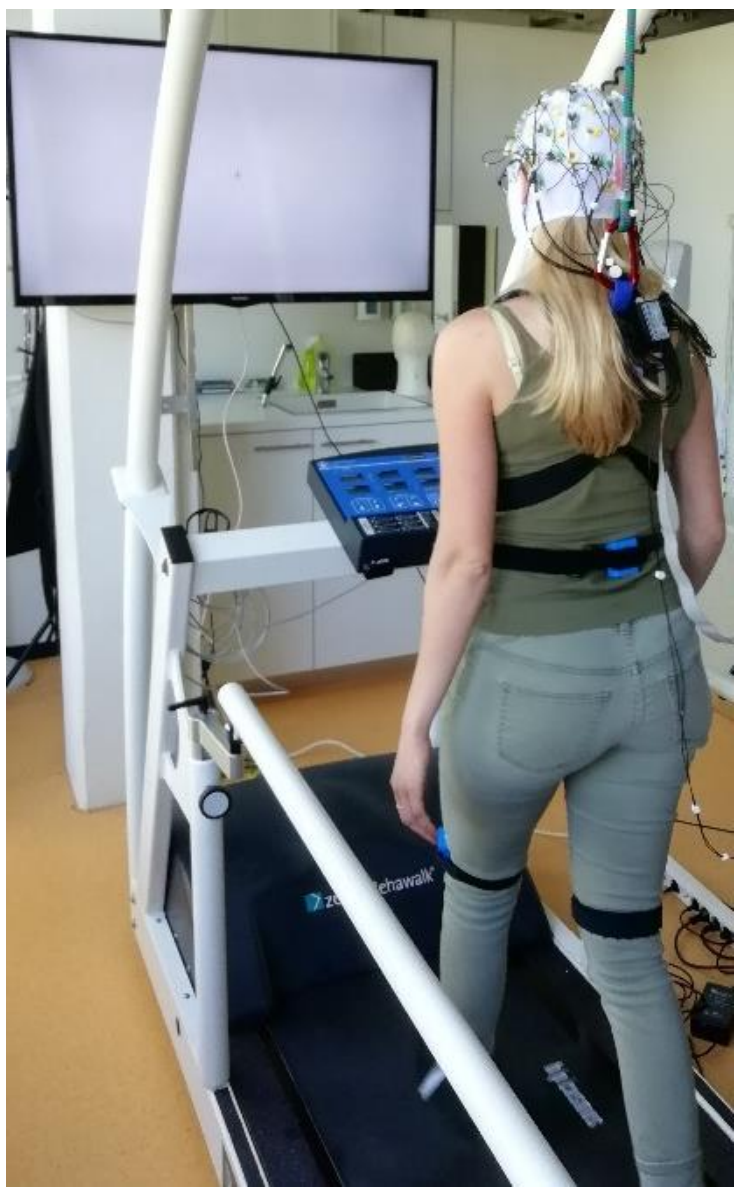

Figure 3: The participant is wearing the 32-channel EEG cap with active electrodes while walking on the treadmill and while gait events are recorded via a force plait implemented in the treadmill. In addition, gait can be tracked via inertial sensors units (IMUs). The inertial sensors are blue and fixed with black straps. Clearly visible are the units at the hip and the thighs. Participants wore no shoes during treadmill walking to avoid heavy concussions. The participants' gait was recorded via the force plate implemented in the treadmill. Picture with permission and copyright of University Ulm, Department of Applied Emotion and Motivation Psychology, Germany.

*3.3 Pressure plate and treadmill*

The pressure plate data is used to record the ground reaction force of the human gait while walking on the treadmill. The pressure plate used is a zebris FDM-T® (zebris Medical GmbH, Isny, Germany) which is integrated into a Pluto® treadmill produced by h/p/cosmos® (h/p/cosmos sports & medical GmbH, Nussdorf-Traunstein, Germany). The raw data was continuously sampled at 100 Hz and the recorded data was later extracted offline. Since it is not yet possible to extract the preprocessed reaction force curves of the feet from the zebris FDM-T recording and analysis software, the recorded data was extracted as raw data and preprocessed offline (see section 3.6 for details).

*3.4 Temporal synchronization unit and data merging*

Since the recording systems (EEG and pressure plate) are independent devices and not coupled and timely synchronized during recording (see structural design in Figure 2), the time synchronization between the two devices has to be realized with a custom technical solution. As shown in Figure 2, the stimulus PC was used for triggering of the recording systems. The stimulus PC runs the software Presentation® (Neurobehavioral Systems) which is capable of sending various trigger output signals via a series of possible ports. In the present study, an LPT port was chosen to connect the EEG system and stimulus PC. This is to avoid jittering of the marker signal, typically occurring with serial ports that exceed the necessary temporal precision for EEG recordings of less than 1-2 milliseconds. The pressure plate (and the IMU system) are triggered directly by the stimulus PC and receive start and stop markers from the stimulation software from Presentation. The trigger interfacing between the stimulus PC and the pressure plate (and, if included, the inertial sensor system) was realized via a custom-made microcontroller-based USB synchronisation device. This device allows to synchronize the recording devices with the existing set-up. Finally, the different recording files need to be written into one output file for data merging and data fusion. The EEG data file was used for data merging and all other data were written to and synchronized with this output file using MATLB® routines.

### *3.4.1 Time synchronization*

The reliability and validity of the results relies on the precision and accuracy with which the data recorded by the different devices and softwares can be temporally synchronized and merged. To minimize the error in the time delay between all recorded datasets, several validation steps were taken into consideration to estimate the different time delays of the different interfaces and the custom made USB communication device. This synchronization device sends the trigger signals from the stimulus PC to the treadmill at the same time the trigger signal is sent to the EEG system. The time delay was explored using the following method. The analog output of the USB synchronisation device comprises a LED. The LED lights up every time an analog trigger signal is created. The LED as well as the PC display are simultaneously recorded with a camera. The accuracy of this method depends on the sampling rate of the camera, which in the present study comprised a frame rate of 120 Hz. The estimation and validation of the time delay for later data synchronization was investigated prior to the pilot study based on test signals. The tests were performed 10 times by triggering the devices in different frequencies. During all tests, the time delay between EEG and treadmill triggers was shown to be below td1 = 8.33 ms and therefore within one time frame of the treadmill (sampling frequency of 120 Hz). There was no jittering. Therefore, it is possible to assign the detected gait events as onset-markers for ERP analysis during the step of data merging and data fusion (see 3.6 for details).

## *3.5 Experimental design and set-up*

The experimental set-up of the pilot study was approved by the ethics committee of Ulm University, Germany (https://www.uni-ulm.de/einrichtungen/ethikkommission-der-universitaet-ulm/). The following inclusion criteria were defined. Healthy young adults (e.g., students from Ulm University, Germany or from Ulm University of Applied Sciences, Germany) with no history of drug abuse, medication, somatic injuries or somatic or physiological or neurological diseases could take part in the pilot study. The participants had to fill in a pre-study-questionnaire to confirm these parameters and give oral and written informed consent for voluntary participation. In total, 3 volunteers (3 females, age: 22-24 years) were included for pilot testing. All participants were wearing glasses, since all four were short-sighted. Participants were asked to walk on the treadmill with bare feet. The experiment was conducted in the EEG-FNIRS-Brain-Imaging Lab of the Department of Applied Emotion and Motivation Psychology at Ulm University, Germany. Further, the interface solutions were tested in the Motion Lab of the Research Group Biomechatronics of Ulm University of Applied Sciences, Germany.

Upon arrival, the participants received detailed instructions about the experiment and were debriefed about its purpose. They were accustomed to the treadmill and the EEG recording (see Figure 3 for an overview). They received practice trials to determine individual preferences for walking on the treadmill at one's own preferred pace. Next, participants were familiarized with the procedure of the experimental-psychological walking paradigm. The paradigm was developed to standardize and control the participants' gait cycle in different conditions of spontaneous and instructed treadmill-walking. In total, seven different experimental conditions were created. These conditions were experimentally designed in a manner such that they manipulated the participants' selective attention to gait. This allowed to produce reliable and reproducible alterations in the gait cycles and to explore gait-related

ERP components as a function of varying degrees of voluntary attention. While walking on the treadmill at the velocity of one's own pace, the participants received detailed written instructions on the TV screen. They were told to walk spontaneously without any specific attention focus (task 1). Next, they received the instruction to take big steps (task 2), small steps (task 3), to concentrate on the left initial contact (task 4) or the right initial contact (task 5), on the left toe-off (task 6) or the right toe-off (task 7). Tasks and walking conditions were randomized and counterbalanced across participants. Each task lasted about 30 seconds and was repeated four times in randomized order. Between each task 30 seconds of normal walking were included. At the start and the end of the experiment, each participant was asked to walk about 4 minutes without any task to have additional data for training of the artificial network (see section 3.9). None of the four participants stated any problems with the tasks given. All participants choose independently to walk with a pace of 3.5 km/h.

### *3.6 Data fusion and data preprocessing*

Data preprocessing and data fusion of the different datasets were performed with MATLAB®. For the pressure plate, preprocessing of the data included the following steps. First, the raw data comprising the pressure measurements of each cell of the plate over time is preprocessed to get the vertical ground reaction force (vGRF) curve for each foot. A segmentation of the pressure measurements into left and right foot is provided by the FDM-T software. Therefore, the vGRF can be calculated by simply summing up corresponding pressure measurements of the left and right foot over time. From this vGRF measurement, it is possible to extract gait-related events. As shown in Figure 4, any of the standardized events of the stance phase of human gait cycle could be detected from the preprocessed pressure plate data, although in our study only initial contact (IC) and toe-off (TO) is used.

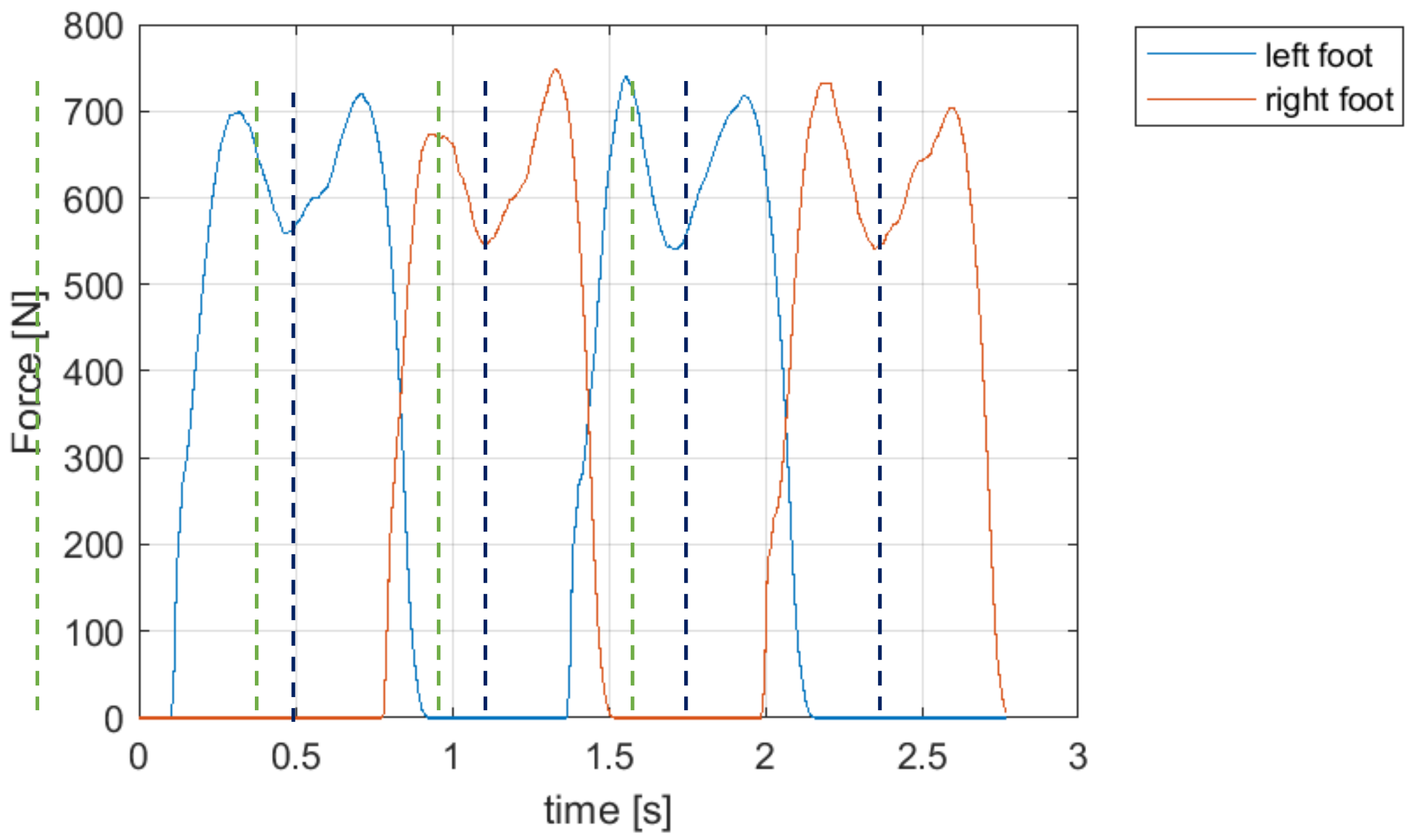

Figure 4: Vertical ground reaction force (vGRF) for left and right foot calculated from pressure plate for gait event extraction. Green vertical lines show the extracted initial contact (IC) events, in black toe-off events (TO).

For the analysis of gait-related ERPs, the (IC) as a prominent gait-event was considered. In total, 3750 (for each person about 1256) events related to the left and right IC could be identified across the seven experimental conditions, resulting in a total of 7500 gait markers for EEG data analysis. The EEGlab toolbox and the Brain Vision Analyzer software were used for EEG data preprocessing and gait-related EEG analysis. The EEG data was filtered from 0.1 - 100 Hz and visually inspected for artifacts. Independent component analysis (ICA) was used for EEG decomposition to separate EEG signals from signals elicited by eye-movements, movement-artifacts and other artifacts. In addition, the data sets were analyzed without using ICA to inspect gait-related ERPs elicited by the left and right IC, respectively.

### *3.7 Data evaluation and marker extraction*

After preprocessing of the EEG data, epochs were extracted from the continuous EEG data using the gait events “IC left” and “IC right” as onset markers. Epochs were extracted from 1000 ms before until 1000 ms after the event (IC) and baseline corrected. The length of the individual epochs (-1000 ms – 1000 ms) was chosen to include the whole gait cycle within one epoch. Two consecutive ICs of the same foot have a time lag of at least 1000 ms, ICs of different feet at least 490ms. As described above, each participant walked with a pace of 3.5 km/h. The epochs were then averaged across all available epochs for each subject suggesting sufficient data quality and good signal to noise ratio. It is principally possible to split the EEG epochs into the different seven walking conditions. It is also possible to analyze the data separately for left and right initial contact (IC). The analysis of gait-related ERPs reported here was first performed for all available epochs, irrespective of the different walking conditions (spontaneous or instruction guided). Next, single analyses were performed for each of the walking conditions.

### *3.8 Data analysis and data classification*

Event-related brain potentials were analyzed from the epochs and the averaged EEG data. Electrodes of interest were chosen in line with the somatotopic representation of movement in the motor cortex and in line with previous EEG research. The results are reported descriptively for each participant and are illustrated in Figure 5 and Figure 6. Moreover, in addition to ERP analysis, ERSP measures were also descriptively analyzed to determine changes in the amplitudes of the EEG frequency spectrum across time (event-related spectral power, ERSP). ERSP maps were analyzed relative to the onset of the initial contact (IC).

### *3.9 Neural networks and data classification*

This setup should later be able to be used for machine learning applications. Those applications can cover but are not limited to the classification of gait phases, detection of gait events and detection of gait disorder based on EEG data. To show the principle applicability, we chose the following classification problem as a proof-of-concept: the network should classify gait phases of single and double support times, i.e. time intervals where only the left or the right foot is on the ground (single support) or both (double support). As the EEG data is highly patient-specific, we trained a neural network for each patient separately. The networks were trained using Mathworks MATLAB® R2019b, Deep Learning Toolbox. We used the fused data from EEG and pressure plate, preprocessed as described above using EEGLab. Only electrodes CP1 and CP2 are selected as input (cf. above). The events of the pressure plate are used as labels, leading to the three classes mentioned above. A total of 1256 IC events are contained in each sequence. A sliding windows of size 1100 samples is moved over the whole EEG data, leading to approx. 114.000 training sequences with the corresponding label information. The size of the sequence resembles the duration of one gait cycle. Afterwards, all windows are shuffled randomly and separated into training (70%), validation (10%) and testing (20%) datasets.

We choose a recurrent deep learning approach because this end-to-end learning strategy requires no manual feature extraction step, which itself is a separate optimisation problem. A simple but effective network architecture based on Long Short-Term Memory Cells (LSTM) [24] is chosen, which is often applied to time sequence classification and regression problems in many different areas [25, 26, 27,].

The network architecture is as follows:

- Input layer of size 2x1100 (according to the number of EEG channels and window size)
- LSTM layer with 50 LSTM units (returning whole sequence)

- LSTM layer with 50 LSTM units (returning only the last value of the sequence)
- Dense layer with 1024 neurons
- Dense layer with 256 neurons
- Dense layer with 3 neurons (according to the number of classes)
- Softmax layer
- Classification layer

Each layer is followed by a dropout layer with dropout probability of 0.2

## 4. RESULTS

### *4.1 Gait-related event-related brain potentials (gERPs)*

Figure 5 (right column) shows the grand mean average of the participants for the averaged IC epochs. Figure 5 (left column) shows the same time plot of a single subject. As can be seen in figure 5, the grand mean average of the gait-related ERP is time-locked to the onset of the initial contact (IC). The event-related potential shows a positive deflection in the time window from -400-0 ms before the start of the IC and an increase in amplitude (negative in voltage) starting immediately at IC onset. This increase in amplitude reaches its maximum across participants at about 100-200 ms post IC onset and lasts for about 400 ms until the start of a second negativity. This ERP modulation, triggered by the IC, is found across subjects (figure 5, right column) and at the single subject level (figure 5, left column). As shown in figure 5, the amplitude of the potential was modulated by instruction, e.g., whether participants attended to the movement, made large or small steps. The potential was most pronounced at centroparietal electrode sites, CP1 and CP2, and observed at central and frontocentral electrodes, as well. Topographic plots show a focus over central and centroparietal electrode sites (C3 and CP1) of the left hemisphere, see figure 6 (top right column). Exploratory source imaging plots comprising the time period from 0-400 ms post IC onset are illustrated in Figure 6. The plots confirm neural activity changes starting in BA 6 (motor cortex) immediately at IC onset, extending to the somatosensory cortex (BA 5) and the inferior parietal cortex (BA 40).

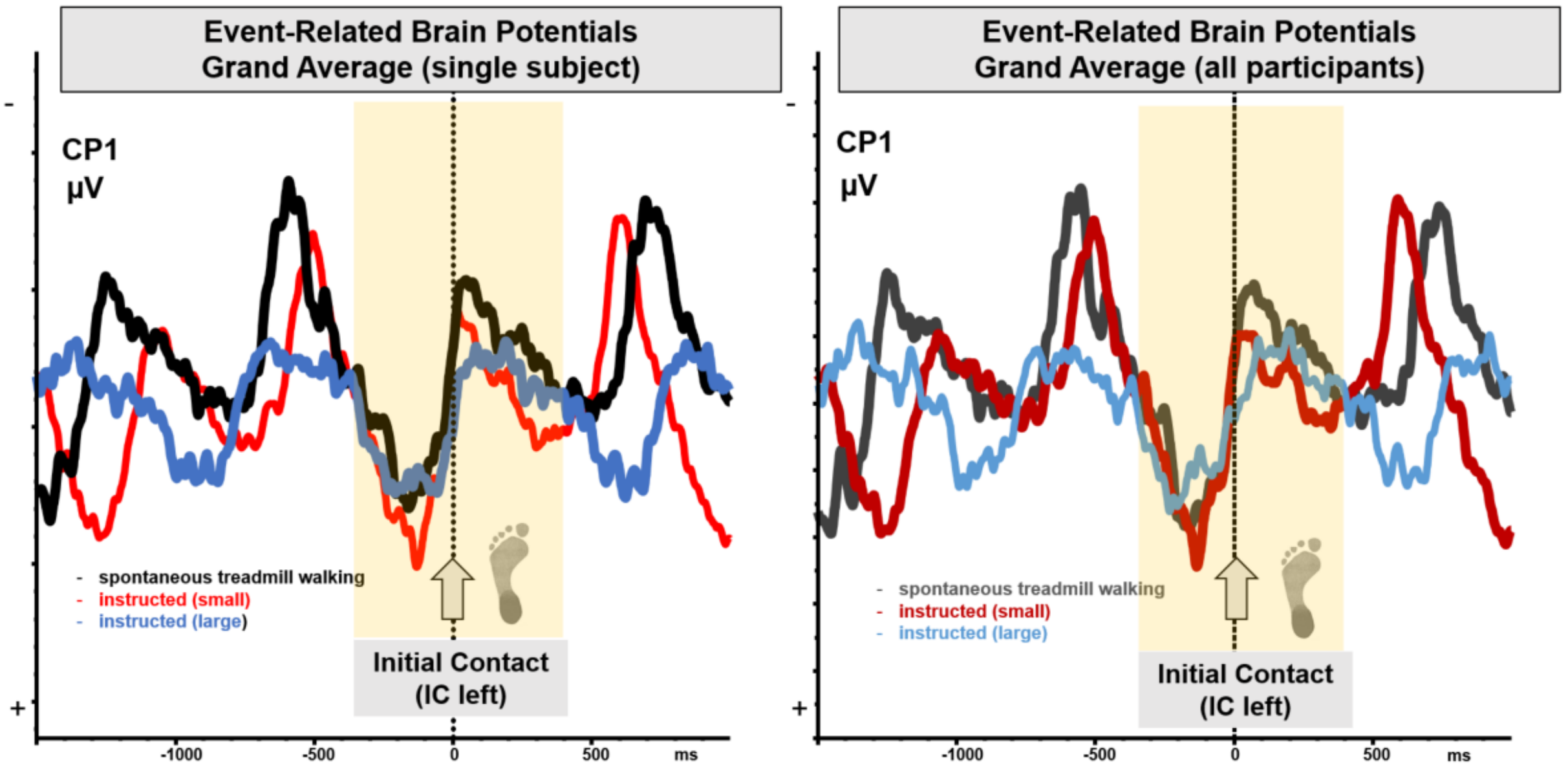

Figure 5: Grand mean ERP waveforms averaged across participants time-locked to the left initial contact of the gait cycle (left column) at the EEG electrode position CP1. Grand average ERP waveforms of a single subject, time-locked to the left initial contact of the gait cycle (right column) at the EEG electrode position CP1.

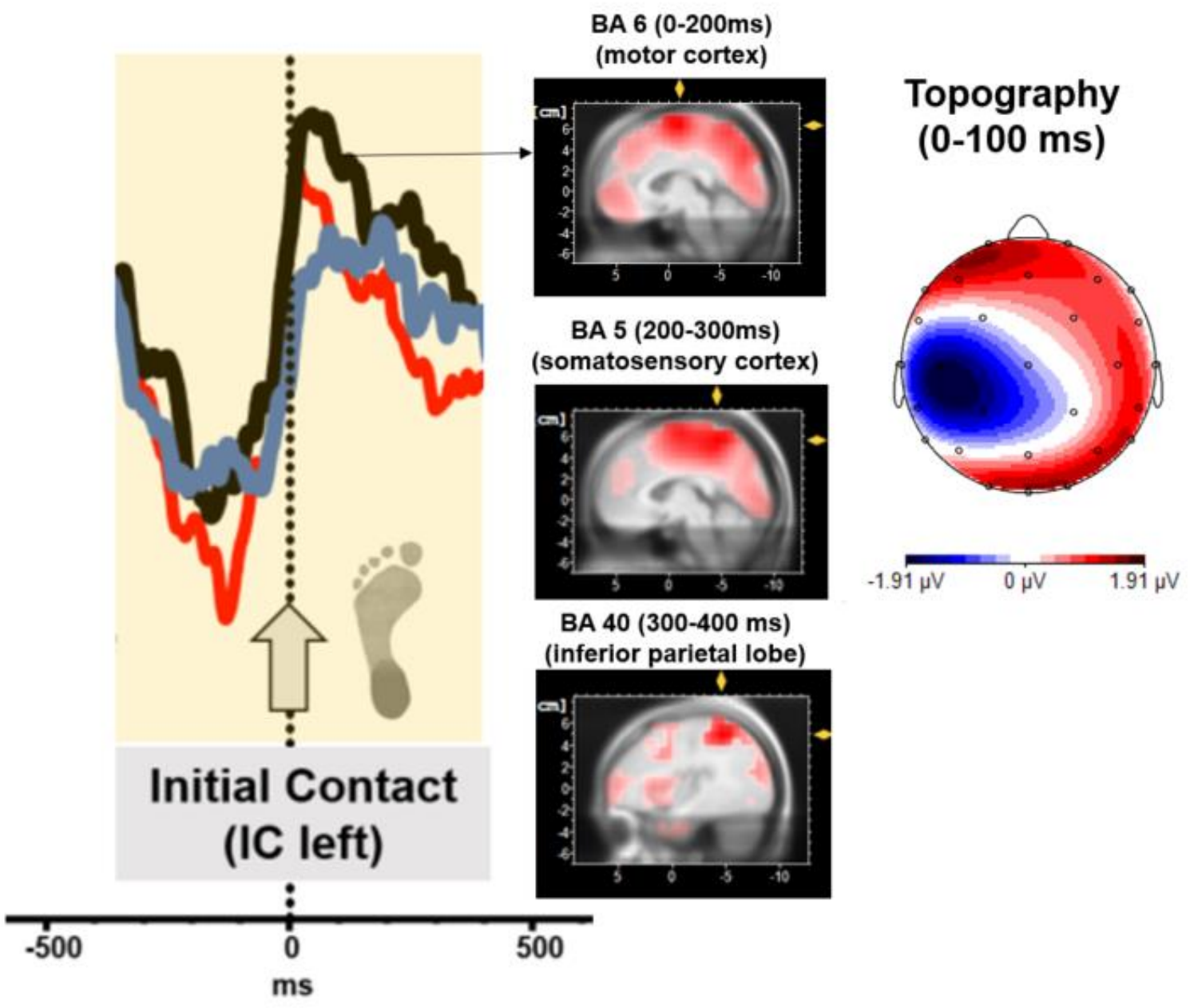

Figure 6: Topography of the motor potential elicited by the initial contact (top right column). Source imaging plots across the time course of the motor potential.

*Frequency analysis (ERSP)*

Further, to explore the neural correlates of the gait cycle in the frequency domain, ERSP plots were calculated with EEGlab toolbox to determine overall changes in frequency patterns across participants. This revealed a change in the frequency power spectrum starting at around 400 ms before IC onset. This change can best be described as a decrease in the alpha band/mu-rhythm (event-related desynchronization), whereas a prominent increase in the alpha power (event-related synchronization) was seen in the time window from 0-300 ms post IC onset. These patterns were observed for the left and right IC events, see figure 7.

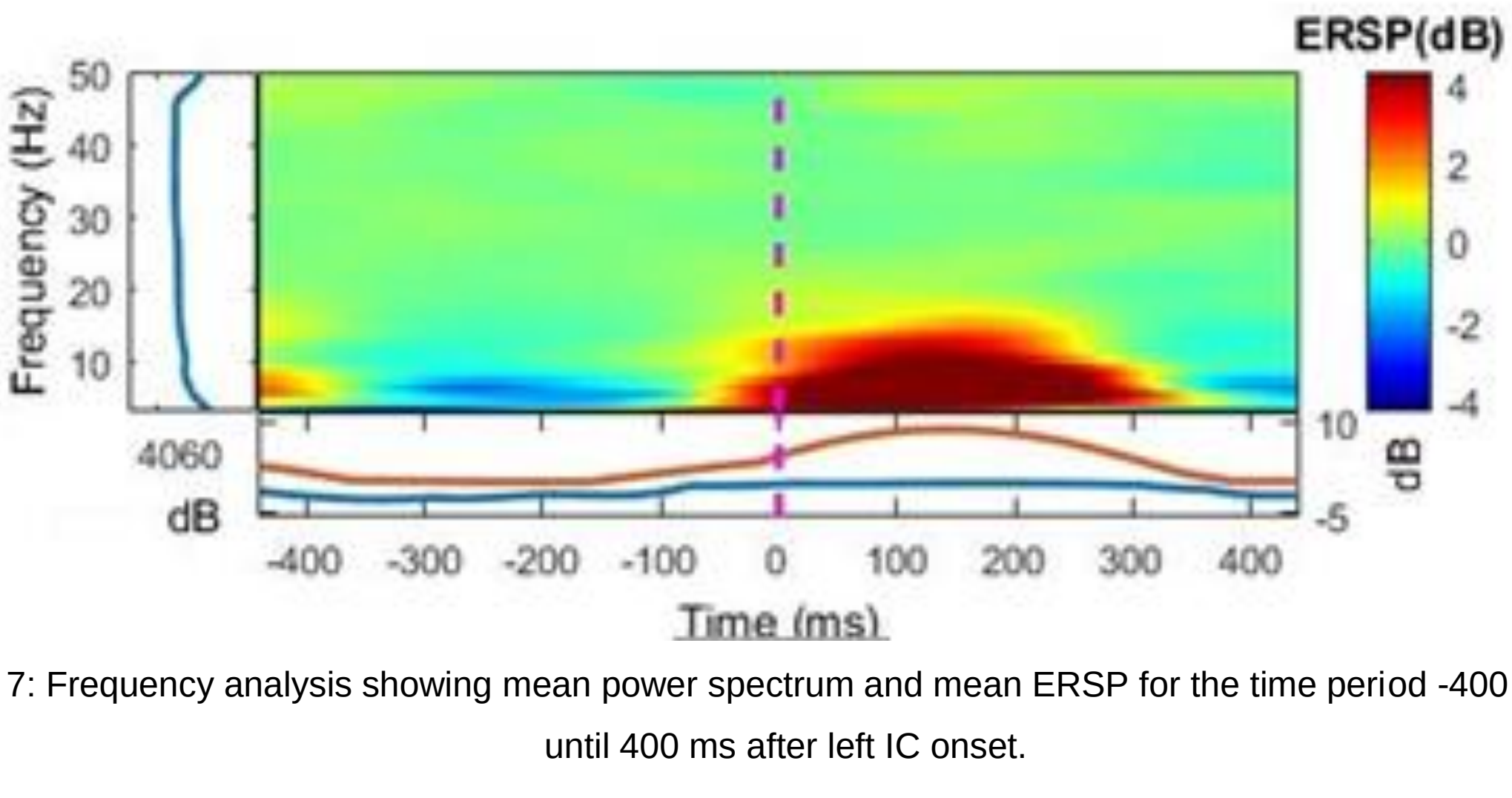

Figure 7: Frequency analysis showing mean power spectrum and mean ERSP for the time period -400 ms before until 400 ms after left IC onset.

*4.2 Neural Network*

The trained gait phase classifier achieved an accuracy of overall 87.4% (subject 1), 89.7% (subject 2) and 83.9% (subject 3) on the test set. The confusion matrices in Figure 8, Figure 9 and Figure 10 show that the network is able to distinguish between left and right single support phases (classes 1 and 2) with a high precision, whereas the separation between single support and double support phases (class 3) has lower performance: class 1 and 2 are confused with class 3 with a higher frequency.

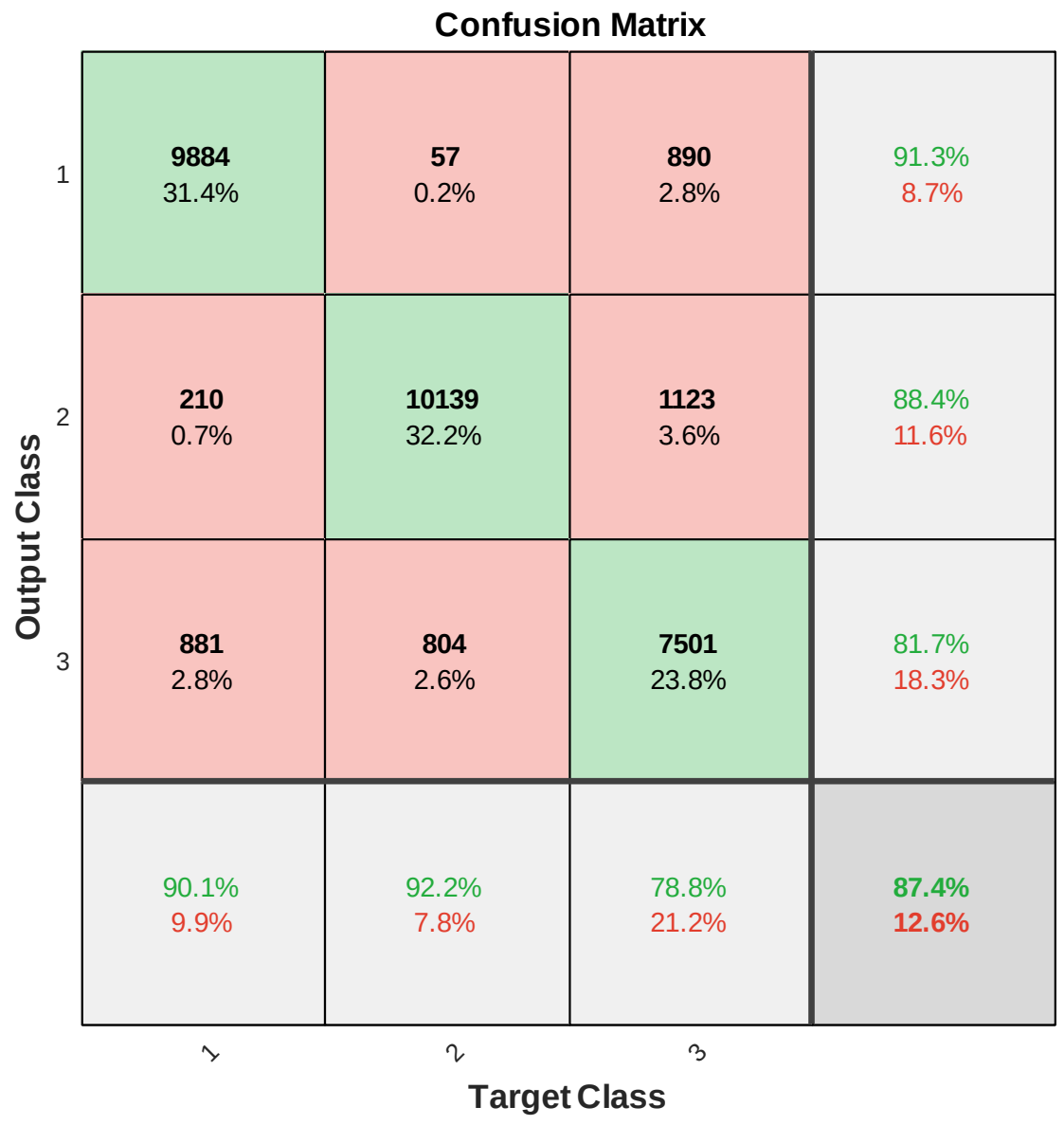

Figure 8: Confusion matrix of classifier for subject 1

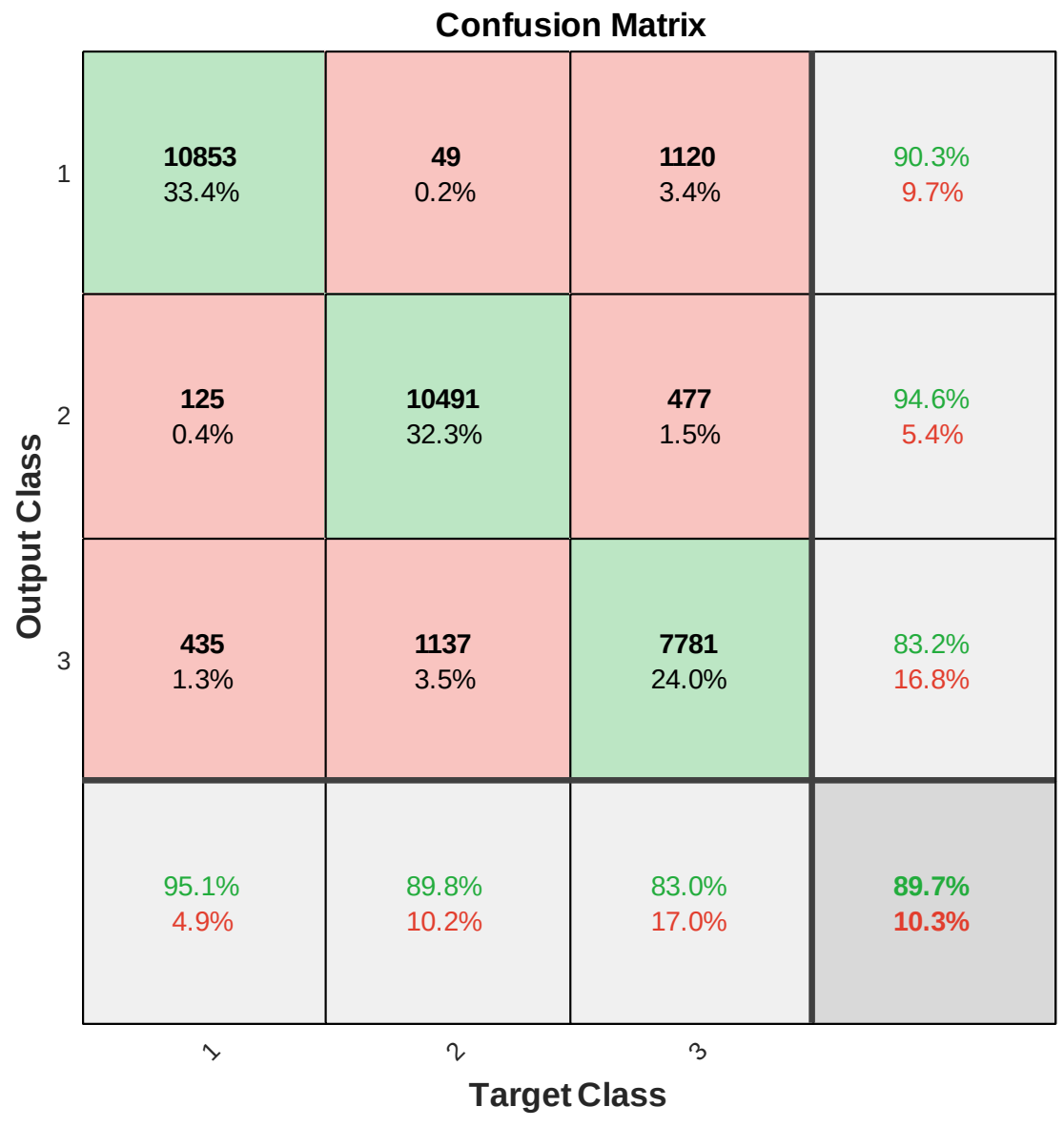

Figure 9: Confusion matrix of classifier for subject 2

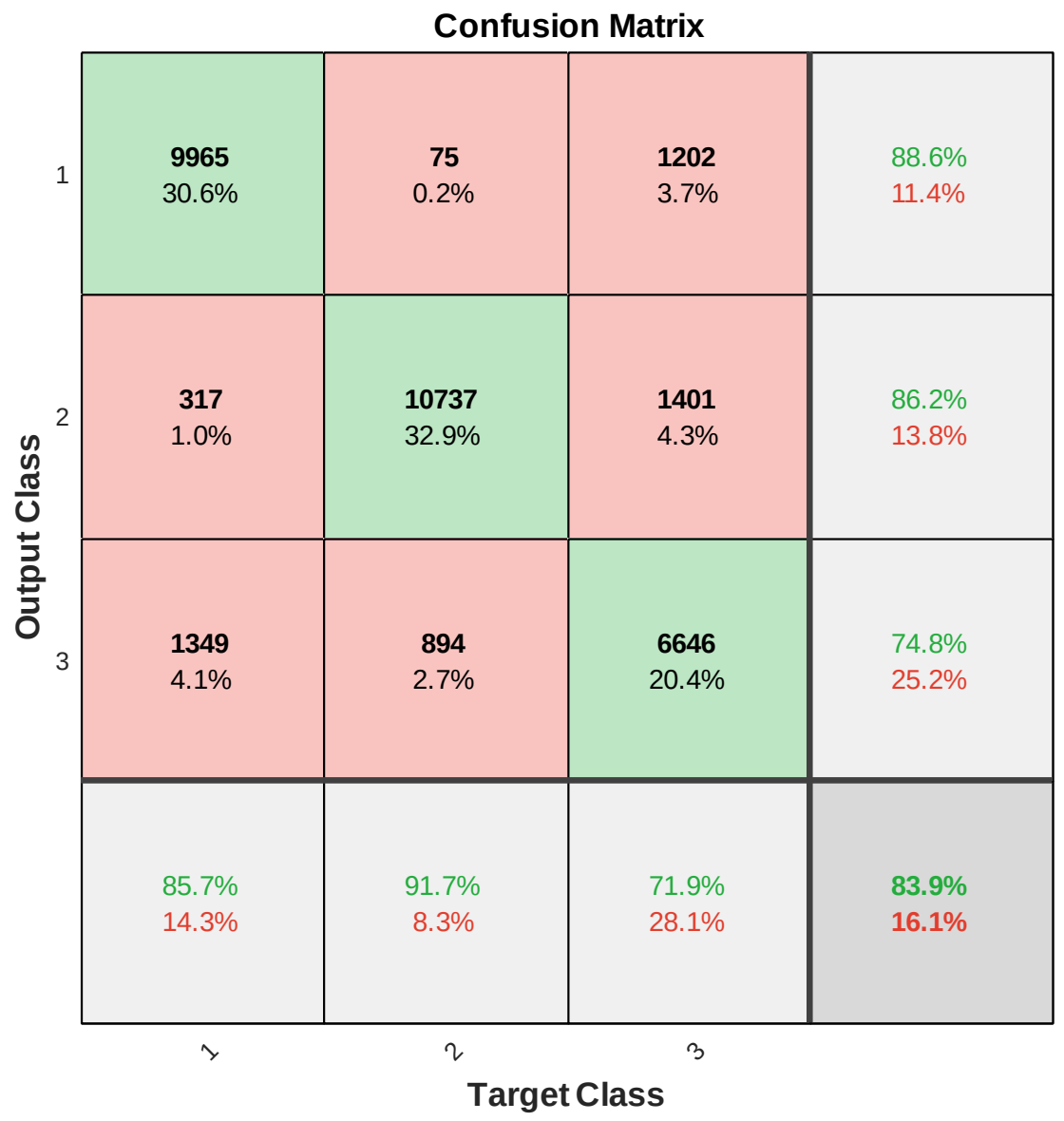

Figure 10: Confusion matrix of classifier for subject 3

## 6. Discussion and Conclusions

This pilot study explored the experimental and technical solutions for the analysis of gait-related brain potentials (gERPs) during treadmill walking based on gait markers taken from pressure plates. Although the attempt to evaluate the neural correlates of human gait by means of EEG is not new, the approach of the present study aimed to investigate open key questions that will now be discussed in detail.

The technical solution investigated in this study allows synchronization of gait- and EEG data from different recording devices and from different recording software installed on different computers. While parallel-ports are the best way to avoid jittering and timely precise triggering, the problem is, that today, these ports are no longer available in customized computers and instead replaced by serial ports or with Bluetooth. These solutions, however, do not reach the accuracy of data transfer of parallel ports. As observed in the validation tests performed in the present study, the custom-made USB synchronization interface - connecting the stimulation unit and the pressure plate - produced a temporal delay of the trigger signal > 2ms. This temporal delay is intolerable for ERP analysis because ERPs are measured time-locked to the onset of the event of interest. Therefore, the data streams recorded need to be synchronized such that the gait-events of interest can be used as reliable makers for EEG epoching and ERP averaging. With the present solution, the custom-build USB interface was developed. The interface controls signal transfer from one device to the other and allows determining the temporal delay elicited during data transfer and using this information for data synchronization. Whereas the first step included a hardware solution, the second step included a software solution using MATLAB as programming platform. Moreover, a parallel port was used for sending triggers from the stimulation unit to the EEG recording unit. These triggers could be used as jitter-free references to which triggers sent from the stimulation unit to the force plate could later on be aligned to.

***Analysis of gait-related event-related brain potentials (gERPs) using gait markers***

As described in detail in the Methods, the individual events of the human gait cycle were derived from data recorded via pressure plates. After preprocessing of the raw data, a prominent gait event, i.e. the initial contact (IC) of the left and right foot could be clearly identified from the recorded data and was used for EEG epoching and ERP averaging. As a gait-event, the IC is characterized by distinct periods of motor preparation and motor execution and therefore, most likely elicits specific motor cortical potentials probably even during spontaneous walking. In the present study, IC related ERP analysis included both, spontaneous and instructed treadmill walking. As shown in Figure 5 and Figure 6, this elicited a prominent increase in negativity, starting immediately after the initial contact (IC) and peaking within the first 100 ms post IC onset. These motor-related changes in cortical activity were observed in individual participants as well as in the grand average waveforms including all participants. The amplitude of the potential was differentially pronounced during spontaneous as compared to e.g., instructed walking. Asking participants for instance to make small or large steps modulated the amplitude of the potential compared to spontaneous walking. This suggests that walking is under the control of cognition and attention [10, 22]. This influence of attention and cognition on gait, is reflected by the modulation of the motor potential elicited by the left and right IC. Topographic plots as well as source imaging plots revealed another interesting finding. Although, the motor potential was most pronounced at left and right frontocentral and centroparietal electrodes, it was topographically most pronounced over left central and centroparietal elecrodes, C3 and CP1. In the 10-20 EEG system, the electrodes C3 and C4 as well as CP1 and CP2 are close to the primary motor and primary somatosensory cortex. Given that the motor potential was found in all walking conditions, this suggests participation of the motor cortex also during spontaneous walking. In addition, gait control seems to be more under the control of the left compared to the right cortical hemisphere. This may have implications for clinical observations of gait asymmetry and brain lesions in stroke patients [28]. Moreover, activity changes extended to the inferior parietal cortex (BA40), forming part of the cortex's fronto-parietal attention network [29].

Taken together, the present results suggest that ERPs of the human gait cycle (such as the IC) can be reliably detected in the EEG recording after temporal synchronization with the gait-event and after standardized preprocessing of the EEG data. Moreover, the use of active electrodes seems helpful to obtain data of sufficient signal to noise ratio. Future studies, using the structural design and experimental set-up of the present study could provide further evidence for gait-related ERPs and their modulation during spontaneous and instructed treadmill walking. As shown in Figure 7 motor potential modulation was also accompanied by a decrease in the alpha/mu-rhythm prior to the IC and an increase in alpha band activity immediately after the onset of the IC. This is in line with other findings suggesting event-related desynchronization (ERDS) and event-related synchronization (ERS) before and after movement execution [16]. Given that so far little is known about the neural correlates of human gait, future studies should include both measures – ERPs and ERDS/ERS to further elucidate cortical control of human gait.

***Implications for the rehabilitation of gait***

The possibility to record ERPs related to human gait in real time offers novel ways for the treatment of people with various gait impairments. The present study suggests that gait-related motor potentials can be reliably identified during treadmill walking by EEG recordings time-locked to the gait-marker derived from gait recording via force plates. Regarding the technological solutions, these analyses can be extended to IMU systems.

***Can gait-related EEG data be trained and classified by an ANN?***

The answer to this question has implications for the development of novel brain-computer interface applications and neurofeedback training. So far, BCIs for movement control are largely based on motor-imagery of leg-, arm- or finger-movements. Such applications are quite helpful, if the aim is to substitute real movements with imagined movements in patients who have lost their limbs, arms etc. In the present study, data was trained by an artificial neural network to discriminate EEG activity by classifying gait phases. A gait phase change corresponds to a toe-off or initial contact event. Although preliminary, it was possible to classify events from non-events with high accuracy. This is very promising, as the chosen architecture is simple compared to other time-series classification approaches. One possible extension could be the application of ConvLSTM networks, i.e. a combination of Convolutional Neural Networks (CNN) and LSTM networks. Those networks have shown to outperform simple LSTM networks due to complex features extracted in the convolutional layers. Besides changes of the architecture, there are number of additional ways of improving the network results, which have not been applied in this work, because the network served as a proof-of-concept of the applicability of the framework.

**Funding**

This research received no external funding.

**Conflict of Interest Statement**

The authors declare no conflict of interest. The research was conducted in the absence of any commercial or financial relationships that could be construed as a potential conflict of interest.